\documentclass[12pt]{article}
\usepackage{amsmath}
\usepackage{graphicx,psfrag,epsf}
\usepackage{enumerate}
\usepackage{natbib}
\usepackage{url} % not crucial - just used below for the URL 
\usepackage{hyperref}
\usepackage{booktabs}
\usepackage{makecell}

\usepackage{todonotes}
\usepackage{xcolor}

\newcommand{\blind}{0}

\begin{document}

\def\spacingset#1{\renewcommand{\baselinestretch}%
{#1}\small\normalsize} \spacingset{1}

%%%%%%%%%%%%%%%%%%%%%%%%%%%%%%%%%%%%%%%%%%%%%%%%%%%%%%%%%%%%%%%%%%%%%%%%%%%%%%

\if0\blind
{
  \title{\bf What makes an Expert? Comparing Problem-solving Practices in Data Science Notebooks}
  \author{Manuel Valle Torre,
    Marcus Specht,
    Catharine Oertel\\
    Delft University of Technology}
% \author{
%     Author 1 \\
%     Group, University\\
%     and \\
%     Author 2 \\
%     Group, University\\
%     }
  \maketitle
} \fi

\if1\blind
{
  \bigskip
  \bigskip
  \bigskip
  \begin{center}
    {\LARGE\bf Title}
\end{center}
  \medskip
} \fi

\bigskip
\begin{abstract}
The development of data science expertise requires tacit, process-oriented skills that are difficult to teach directly. 
This study addresses the resulting challenge of empirically understanding how the problem-solving processes of experts and novices differ. 
We apply a multi-level sequence analysis to 440 Jupyter notebooks from a public dataset, mapping low-level coding actions to higher-level problem-solving practices. 
Our findings reveal that experts do not follow fundamentally different transitions between data science phases than novices (e.g., Data Import, EDA, Model Training, Visualization). 
Instead, expertise is distinguished by the overall workflow structure from a problem-solving perspective and cell-level, fine-grained action patterns.
Novices tend to follow long, linear processes, whereas experts employ shorter, more iterative strategies enacted through efficient, context-specific action sequences. 
These results provide data science educators with empirical insights for curriculum design and assessment, shifting the focus from final products toward the development of the flexible, iterative thinking that defines expertise—a priority in a field increasingly shaped by AI tools.
% \textcolor{red}{Maybe we can add an image, a conceptual representation of the sequences or something - Catha}
\end{abstract}

\noindent%
\vfill

\newpage
\spacingset{1.45} % DON'T change the spacing!

\section{Introduction}
Data Science Education (DSE) is widely recognized as a complex, interdisciplinary field that requires more than just technical proficiency (\cite{hazzan2023GuideTeachingData}). 
The consensus in the literature is that DSE should focus on cultivating creative problem-solving skills, prioritizing a learner's ability to navigate ambiguity over rote memorization of tools and algorithms (\cite{donoghue2021TeachingCreativePractical}).
Consequently, data science is increasingly taught as a process—a structured yet flexible workflow that guides practitioners from problem definition to final solution (\cite{schwab-mccoy2021DataScience2020}).

While we teach data science as a process, we lack a deep, empirical understanding of how experts and learners actually enact this process (\cite{ramasamy_workflow_2022}).
The crucial procedural skills involved are often tacit—second nature to experts but opaque and misunderstood by novices (\cite{nokes2010ProblemSolvingHumana}). 
This makes them difficult to observe and, therefore, difficult to teach directly (\cite{anderson1993ProblemSolvingLearning}). 
The recent rise of Generative AI tools only increases the urgency of this problem. 
While powerful, these tools can encourage a superficial, task-completion approach, potentially allowing learners to offload critical thinking and reinforcing the very novice-like behaviors we seek to overcome (\cite{fan_beware_2024}). 
Understanding and supporting the process of developing expertise is therefore more critical than ever (\cite{molenaar_temporal_2022}).

This study addresses that gap by using computational methods to make these learning processes visible. 
We apply a range of sequence analysis techniques, guided by a multi-level framework for educational research (\cite{valletorre2024SequenceMattersLearning}), to investigate the problem-solving process in a large dataset of Jupyter notebooks. 
By doing so, we aim to answer the following research questions:
\begin{itemize}
    \item What characteristic problem-solving strategies can be identified in data science notebooks at different granularity levels?
    \item At which of these levels, if any, can these strategies effectively differentiate novices from experts?
\end{itemize}

The findings from this work will provide a deeper understanding of how data science expertise develops. 
This knowledge can be used to inform pedagogical practices, design more effective learning materials, and develop intelligent support systems that help learners develop the flexible, adaptive skills of an expert practitioner, allowing for better human-AI collaboration in the future (\cite{zhao2023DataMakesBetter,lopez-pernas2021PuttingItAll}).

\section{Related Work}
\subsection{Analyzing the Data Science Process}
Teaching and learning data science is challenging, as it requires integrating statistics, programming, and domain-specific problem-solving (\cite{posner2024CrossdisciplinaryReviewIntroductory, hazzan2023GuideTeachingData,donoghue2021TeachingCreativePractical}). 
In education, this is often structured around a well-defined workflow, such as the Cross-Industry Standard Process for Data Mining (CRISP-DM), which outlines key phases like data preparation, modeling, and evaluation (\cite{martinez-plumed2021CRISPDMTwentyYears}). 
This workflow provides a necessary scaffold for novices (\cite{hazzan2023GuideTeachingData}). 
However, a key challenge in Data Science Education (DSE) is moving learners beyond a rigid, step-by-step execution of this workflow toward the more fluid, iterative, and adaptive process that characterizes expert practice (\cite{ramasamy_workflow_2022}). 

The recent introduction of Generative AI tools only magnifies this challenge.
These tools can automate code writing and debugging, assist with explanations and examples, and generate learning materials for data science (\cite{ellis2023NewEraLearning}).
However, research suggests they can also promote a dependency that stifles self-regulation and encourages a superficial, task-completion approach, potentially leading to 'metacognitive laziness', as described by \cite{fan_beware_2024}.

This makes it critical to shift the focus from the outcomes to understand and support learning as a process (\cite{molenaar_temporal_2022}). 
To do so, researchers need not only authentic data from real-world tasks, but data that is also rich enough to capture the sequence of actions that constitute that process (\cite{saqr2023TemporalDynamicsOnline}).
Computational notebooks, such as those from platforms like Kaggle, offer a rich source of such data, capturing the sequenced actions of thousands of practitioners solving complex, real-world problems (\cite{pimentel2019LargeScaleStudyQuality, quaranta2021KGTorrentDatasetPython}). 
Prior educational research using notebook data has focused primarily on the final product or static code quality, analyzing, for example, cell-level mistakes by \cite{singh2024InvestigatingStudentMistakes}, overall code metrics by \cite{nguyen2021ExploringMetricsAnalysis}, or documentation quality by \cite{wang2021WhatMakesWellDocumented}. 
While valuable, these approaches overlook the dynamic, sequential nature of the work (\cite{valletorre2024SequenceMattersLearning}). 
They can tell us what was produced, but not how the author arrived at the solution (\cite{knight2017TimeChangeWhy}). 
The critical gap in the literature, therefore, is an understanding of the sequential process itself—the strategies and habits that differentiate novice and expert problem-solvers (\cite{nokes2010ProblemSolvingHumana, zhao2023DataMakesBetter, ramasamy_workflow_2022}).

\subsection{A Multi-Level Framework for Data Science Process Analysis}
\label{sec:relatedFramework}
To analyze these complex workflows, we adopt the principled structure outlined by \cite{valletorre2024SequenceMattersLearning} in their five-component framework for educational sequence analysis. 
This guides our work by ensuring a clear purpose, grounding in theory, and appropriate selection of techniques. 
Following this structure, we first establish a theoretical lens through the literature of problem-solving, which treats complex tasks as a series of cognitive and metacognitive practices (\cite{wang2021AutomatingAssessmentProblemsolving}). 
Specifically, we adopt the five-stage model of authentic problem-solving practices identified by \cite{wang2023ApplyingLogData}:
\begin{enumerate}
    \item Problem definition and decomposition: Practices to understand and simplify a problem, such as breaking it down into smaller subproblems.
    \item Data collection: Actions and decision-making to collect the data needed to solve the problem.
    \item Data recording: How problem solvers keep track of the data they have collected.
    \item Data interpretation: Applying domain knowledge to make sense of collected data and reach a solution.
    \item Reflection: Processes to monitor progress and evaluate the quality of a solution.
\end{enumerate}
This framework allows us to abstract the technical data science actions into a higher-level, cognitively-grounded process, providing a perspective to understand the overall strategic shape of a workflow.

While this high-level framework provides an essential starting point, expertise often manifests at different levels of granularity. 
An expert might follow the same high-level stages as a novice but execute the steps within those stages with far greater efficiency. 
A single analytical method is often insufficient to capture this complexity (\cite{samuelsen2019IntegratingMultipleData}). 
Therefore, this study adopts a multi-level sequence analysis approach to investigate the data science process at different resolutions:
\begin{itemize}
    \item Overall Process Structure: Analyzing the sequence of broad problem-solving stages to identify high-level strategic archetypes.
    \item Phase-Level Transitions: Modeling the probabilistic transitions between specific DS phases (e.g., from EDA to Model Training) to understand the workflow's order of operations.
    \item Fine-Grained Action Patterns: Discovering recurring, low-level sequences of actions within a given problem-solving stage (e.g., the specific steps taken during Data Collection).
\end{itemize}
This multi-level approach, which has been used to understand complex learning in other domains (\cite{matcha2020AnalyticsLearningStrategiesa}), is essential for deconstructing the data science process. 
It allows us to investigate not only what stages practitioners follow, but also how they transition between them and what specific actions they take to enact their strategies. 
This study applies this approach to investigate the differences between the processes of novices and experts (\cite{anderson1993ProblemSolvingLearning}), addressing a critical need to better understand and teach the process of data science. 
Such an understanding is a prerequisite for designing learning tasks and support systems able to assist learners in regulating their own work, especially when using modern AI tools (\cite{xu_enhancing_2025}).

\section{Dataset and Methods}
\subsection{Dataset} %%%%%%%%%%%%%%%%%%%%%%%%%%%%%
\label{sec:dataset}
% Our study uses Code4ML \cite{drozdova2023Code4MLLargescaleDataset}, a set of labelled notebooks from Kaggle\footnote{\href{https://www.kaggle.com/}{https://www.kaggle.com/}}, to analyze patterns in Data Science (DS) by treating DS challenges as problem-solving tasks.

Our study uses Code4ML by \cite{drozdova2023Code4MLLargescaleDataset}, a set of labelled notebooks from Kaggle, to analyze patterns in Data Science (DS) by treating DS challenges as problem-solving tasks. 
We focus on a single Kaggle competition to minimize variability caused by different data modalities (e.g., image recognition vs. text processing). 
We selected the competition with the most notebooks available in Code4ML: the \textit{Quora Insincere Questions Classification}. 
This competition was chosen because it involves a common data science task (text classification), provides a sufficient number of notebooks for analysis, and requires a range of typical data science activities, including data cleaning, feature engineering, classification, visualization, and evaluation (\cite{chakraborty2023QuoraInsincereQuestions}).

To mitigate the impact of outliers, we removed the top and bottom 5\% of the notebooks based on length (\cite{matcha2020AnalyticsLearningStrategiesa}), resulting in a dataset of 440 notebooks. 
The classification of participants into novice and expert categories is based on their Kaggle progression tier. 
While imperfect, these tiers are based on provable metrics and achievements (competition medals, votes by experts, etc.)  \footnote{\href{https://www.kaggle.com/progression}{https://www.kaggle.com/progression}}.
Novices are defined as users in the \textit{Novice or Contributor} tiers (n=339, M length=20.3 cells, SD=10.2), while experts are those in the \textit{Expert, Master, or Grandmaster} tiers (n=101, M length=16.7 cells, SD=6.6). 
Data preprocessing was performed using \texttt{Python}, integrating files from Code4ML and the KGTorrent database to filter the corresponding authors, notebooks, and cell labels (\cite{drozdova2023Code4MLLargescaleDataset, quaranta2021KGTorrentDatasetPython}).

\subsection{Methodology} %%%%%%%%%%%%%%%%%%%%%%%%%%%%%%%%%%%%%%
%%%%%%%%%%%%%%%%%%%%%%%%%%%%%%%%%%%%%%%%%%%%%%%%%%%%%%%%%%%%%%%
We follow the five-component framework for applying sequence analysis in education by \cite{valletorre2024SequenceMattersLearning}: defining the purpose, identifying units and scope, grounding the analysis in theory, selecting appropriate techniques, and interpreting the results for educational intervention.

The \textbf{purpose} of our analysis is to identify discernible problem-solving strategies within the notebook data and determine if these strategies can differentiate between novice and expert data scientists. 
This aligns with our overarching goal of understanding how expertise develops in data science to better inform pedagogy.

For the \textbf{units} and \textbf{scope} of analysis, we treat each labelled code cell as a single action and each Jupyter notebook as a complete, self-contained problem-solving sequence.

\begin{table}
\scriptsize
\caption{Mapping Between Problem-Solving Practices and Code4ML Data Science Phase}
\label{tab:mapping}
\begin{tabular}{l p{2.7cm} p{9.5cm}}
\toprule
% The header row is now using \makecell for consistent alignment and manual line breaks
\textbf{\makecell[l]{Practice}} & \textbf{\makecell[l]{DS Phase}} & \textbf{\makecell[l]{Explanation}} \\
\midrule
\makecell[l]{Problem Definition \\ and Decomposition} 
& Environment & Setting up the environment (e.g., \texttt{install} and \texttt{import modules}) represents an initial decomposition of the problem into a set of solvable tasks that will be handled by specific libraries. \\
& Data Extraction & Extracting data from various sources implies understanding the problem's requirements. \\
& [Data] Debug & Debugging the data involves identifying and correcting errors, which requires understanding the problem and the expected data characteristics. \\
\midrule
\makecell[l]{Data \\ Collection} 
& Data Transform & Data transformation practices, such as \texttt{feature engineering}, \texttt{drop columns}, or \texttt{merge}, are direct actions to shape the raw data into the necessary clean and relevant dataset required to solve the problem. \\
& Exploratory Data Analysis & Understanding patterns and characteristics of a dataset, with \texttt{count unique} or \texttt{count duplicate}, is a crucial part of ``collecting data to solve the problem". \\
& Model Training & Actions within training, like \texttt{choose model class} and \texttt{predict on train}, can be seen as ``actions to collect data," for example, about the relationships between variables. \\
\midrule
\makecell[l]{Data \\ Recording} 
& Data Visualization & Creating a visualization (\texttt{plot metrics}, \texttt{heatmap}, \texttt{distribution}) produces a persistent artifact that records and summarizes the state of the data or model at a specific point, allowing the practitioner to track their findings. \\
\midrule
\makecell[l]{Data \\ Interpretation} 
& Model Interpretation & Actions such as \texttt{print shapley coeffs} and \texttt{features selection} are explicit methods for understanding what the model has learned and using that understanding to refine the solution. \\
& Hyperparameter Tuning & Tuning hyperparameters, by defining \texttt{search model} or \texttt{search space}, is guided by the interpretation of the model's performance on the data. \\
& Model Evaluation & Evaluating model performance is a key step in interpreting the model's effectiveness and making sense of the data. \\
\midrule
Reflection 
& Data Export & This encompasses the process of ``evaluating the quality of the solution." Actions like \texttt{save to csv} or \texttt{prepare output} often represent a final assessment and synthesis of the findings, signifying the end of a cycle and reflecting a judgment on the result. \\
\bottomrule
\end{tabular}
\end{table}
Our analysis is guided by the \textbf{educational principle} of data science as a form of problem-solving (\cite{donoghue2021TeachingCreativePractical,anderson1993ProblemSolvingLearning}).
We use the framework from \cite{wang2023ApplyingLogData} to map the 11 technical DS phases from Code4ML to five general problem-solving practices, based on the actions within each phase from \cite{drozdova2023Code4MLLargescaleDataset}. 
The mapping is detailed in Table \ref{tab:mapping}.
While conventional data science process models like CRISP-DM are useful for outlining an idealized workflow, they are less effective at capturing the enacted \textit{cognitive process} of problem-solving. 
By adopting this cognitive lens, we can interpret actions like `model training' as a crucial `data collection' practice where the practitioner is actively gathering information about feature effectiveness. 
This perspective is essential for revealing the underlying problem-solving strategies that a standard workflow model would otherwise obscure.

%%%%%%%%%%%%%%%%%%%%
For the \textbf{analysis}, we employ a suite of sequence analysis techniques to investigate the notebooks at different levels of granularity. 
This multi-level approach allows us to build a comprehensive picture of the problem-solving processes.
\begin{enumerate}
    \item \textbf{Overall Process Structure:} To identify high-level workflow patterns, we use Agglomerative Hierarchical Clustering (AHC) on the full sequences of problem-solving practices. 
    We use Optimal Matching (OM) to calculate the dissimilarity between sequences and Ward's method for clustering (\cite{saqr2024SequenceAnalysisEducation}). 
    This approach groups notebooks that share a common overall structure or 'narrative'(\cite{studer2016WhatMattersDifferences}).
    \item \textbf{Phase-Level Transitions:} To model the entire workflow as a process and examine transitions between phases, we use Process Mining (PM) and Markov Models (MM).
    \textit{Process Mining} provides a holistic view of the most common pathways through the DS workflow, assuming the author is following a plan (\cite{bannert2014ProcessMiningTechniquesa}).
    We specifically use the technique of \textit{Process Discovery} on the \texttt{bupaR} package in \texttt{R}, which provides similar results to Fuzzy Miner, the most popular algorithm for Educational Process Mining (\cite{saint2022TemporallyfocusedAnalyticsSelfregulated}).
    To ensure clarity, process maps visualize the top 80\% most frequent transitions.
    \textit{Markov Models}, using \texttt{seqHMM} in \texttt{R}, focus on the probabilistic transitions between states (i.e., DS phases), allowing us to find differences between Novices and Experts based on their transition profiles (\cite{helske2024ModernApproachTransition}).
    Furthermore, we can use the initial models and the Expectation Maximization algorithm to cluster the sequences, potentially finding major representations of Experts or Novices per cluster.
    \item \textbf{Within-Phase Action Patterns:} To find common, low-level action sequences within a specific problem-solving phase, we use Sequential Pattern Mining (SPM). 
    Using \texttt{PrefixSpan} in Python, we search for frequently occurring sub-sequences of specific DS actions (e.g., import\_modules, count\_missing, save\_to\_csv) per problem-solving stage. 
    This is ideal for discovering fine-grained procedures in the specific context of each stage (\cite{yan2020AutoSuggestLearningtoRecommendData}). 
    We set a minimum support threshold of 5\% due to the density of information in sequences within stages (\cite{akhuseyinoglu2022ExploringBehavioralPatterns}).
\end{enumerate}
Following the clustering analyses (AHC and MM), we use a Chi-square test of independence to assess whether cluster membership is significantly associated with the author's expertise level.

Finally, while designing a specific \textbf{educational intervention} is beyond this study's scope, we discuss the implications of our findings for DSE curriculum, teaching practices, and the design of support tools in the Discussion section (\ref{sec:discussion}.

\section{Results} %%%%%%%%%%%%%%%%%%%%%%%%%%%%%%%%%%%%%%%%%%%%%%%
%%%%%%%%%%%%%%%%%%%%%%%%%%%%%%%%%%%%%%%%%%%%%%%%%%%%%%%%%%%%%%%%%
\label{sec:results}
This section presents the results of our sequence analysis, structured according to the multi-level framework described in the Methodology. 
For each level of granularity—overall process structure, phase-level transitions, and within-phase actions—we first identify characteristic patterns and strategies (RQ1) and then assess whether these patterns differentiate authors by their expertise level (RQ2).
\begin{figure}
    \includegraphics[width=0.9\textwidth]{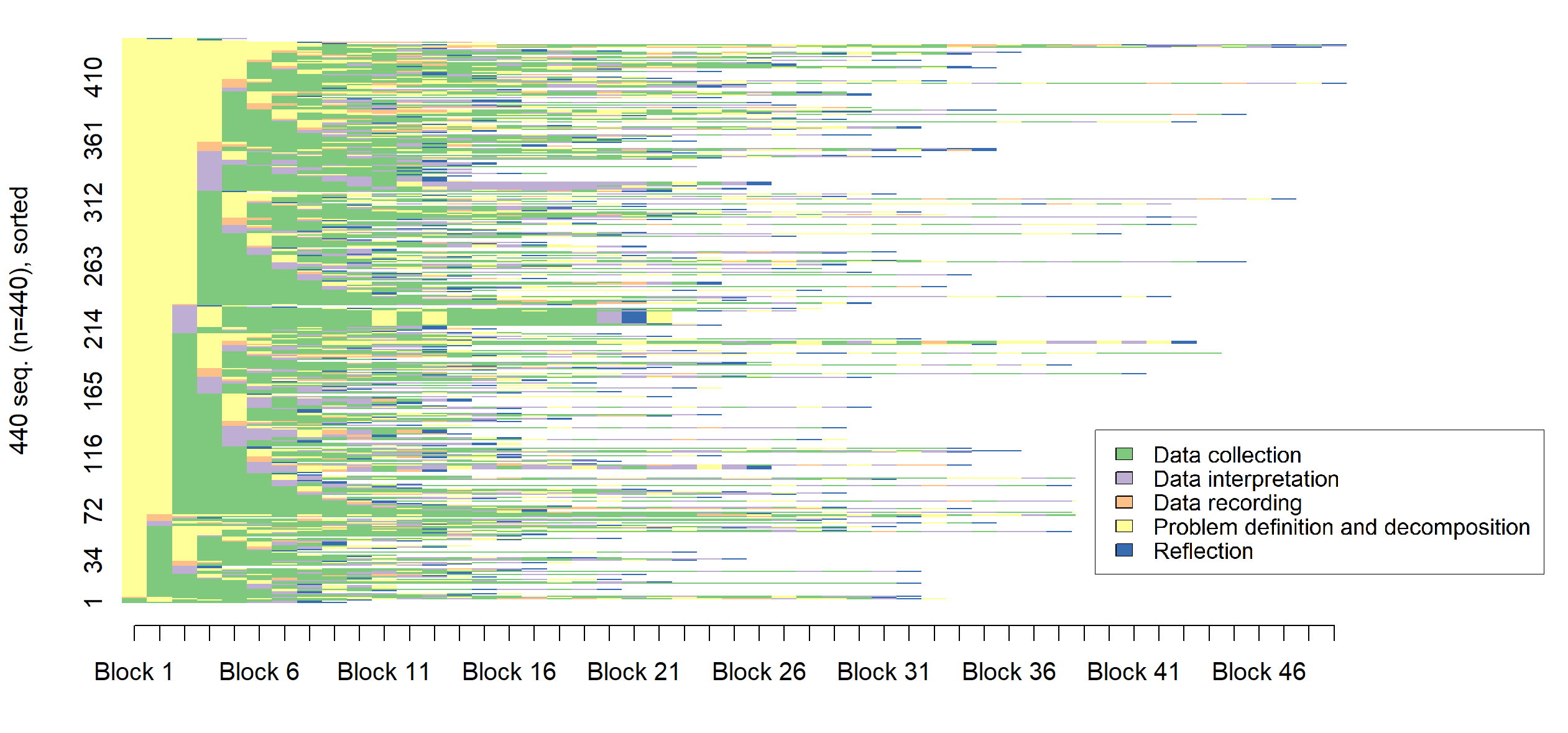}
    \caption{Visualization of all Problem-solving Sequences}
    \label{fig:problemSequences}
\end{figure}
\subsection{Exploratory Analysis and Overall Process Structure}
\begin{figure}
    \centering
    \includegraphics[width=0.9\linewidth]{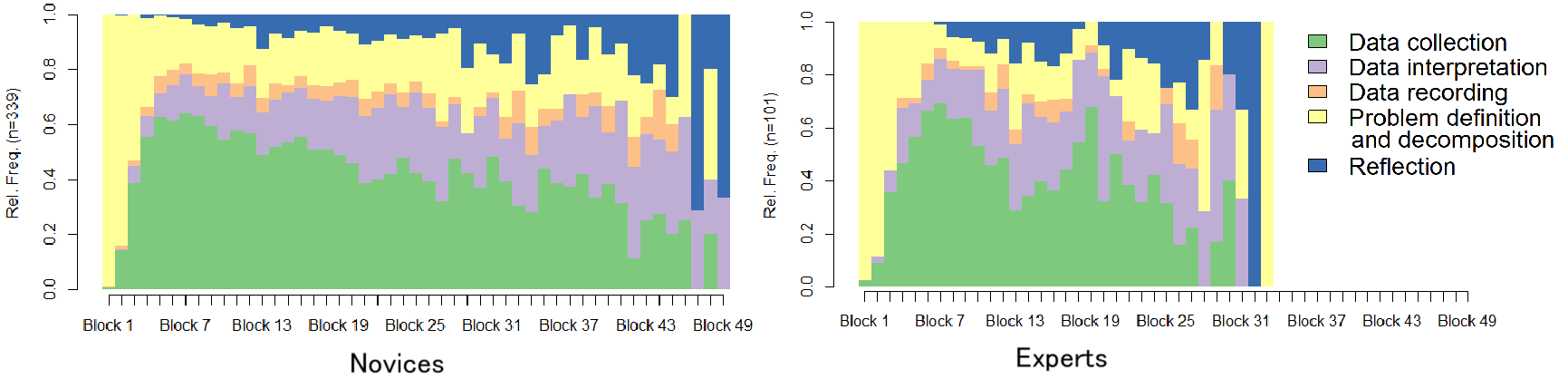}
    \caption{Relative Frequency visualization of Novice and Expert Sequences}
    \label{fig:densityNoviceExpert}
\end{figure}
\begin{figure}
    \centering
    \includegraphics[width=0.7\linewidth]{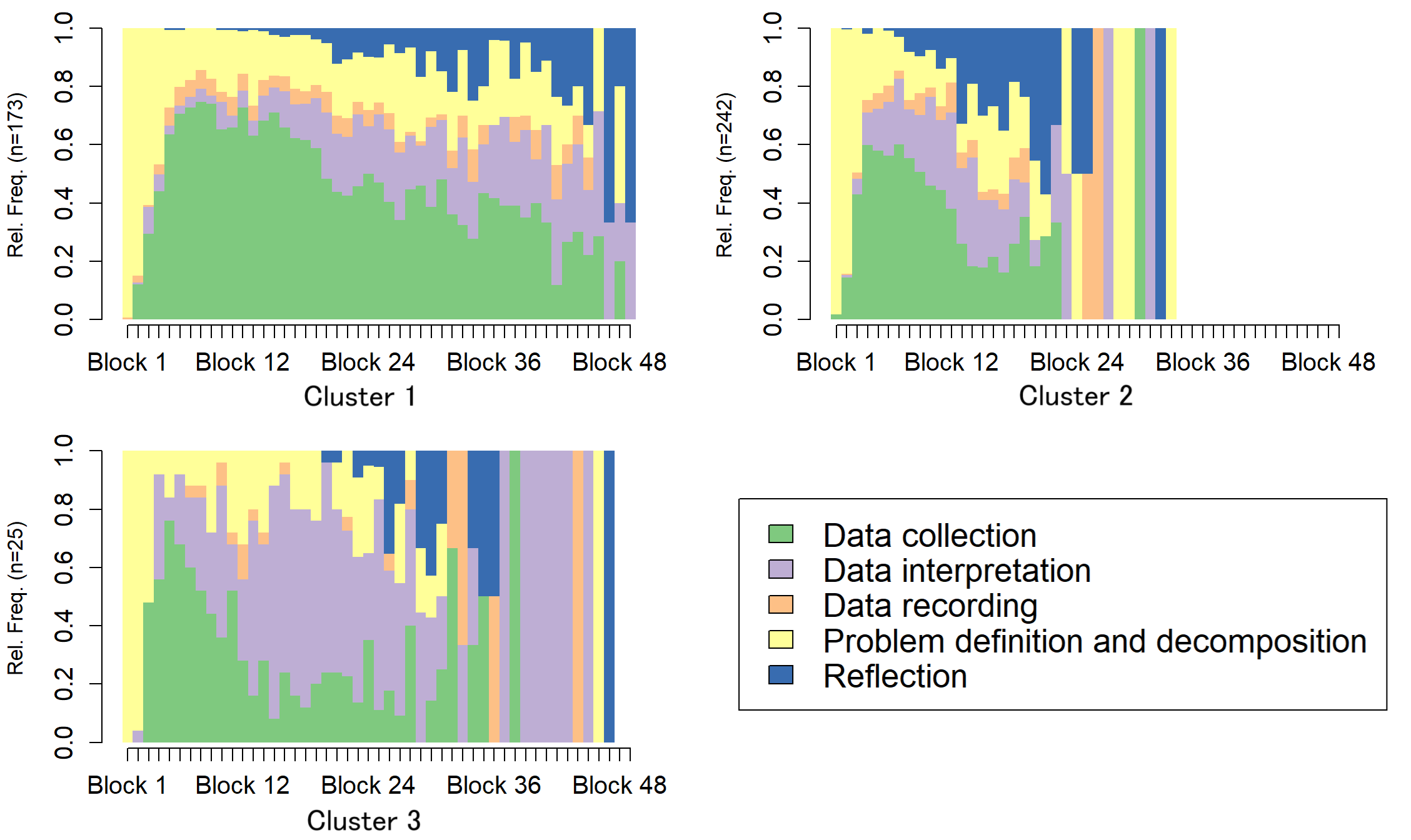}
    \caption{Relative Frequency visualization of Clustered Sequences}
    \label{fig:densityClusters}
\end{figure}
A visual inspection of all 440 sequences mapped to problem-solving stages, in Figure \ref{fig:problemSequences}, reveals common structural characteristics (RQ1). 
Most sequences begin with one to three \textit{Problem Definition and Decomposition} actions (yellow) and conclude with \textit{Reflection} (blue). 
The phase of \textit{Data Collection} (green) is prevalent throughout the process. 
Beyond this initial structure, the sequences show high variability, suggesting multiple pathways through the problem.

The relative frequency of actions over time (Figure \ref{fig:densityNoviceExpert}) shows that novice notebooks tend to be longer than expert notebooks.
Furthermore, experts exhibit a second peak of \textit{Data Collection} activity later in their process (around step 19), whereas for novices, \textit{Data Collection} peaks early and then steadily declines.
% This likely represents an iterative process where experts go back to feature engineering and model training after an initial \textit{Data Intepretation} (in lilac), while novices closely follow general problem-solving heuristics \cite{nokes2010ProblemSolvingHumana}.

% \textcolor{red}{The differences in sequence length and the timing of the Data Collection peak suggest that novices might be engaging in a more linear, step-by-step process, while experts are iterating more between data collection, analysis, and interpretation.
% This aligns with the literature on expert-novice differences in problem-solving, where experts are known to iteratively re-assess problems as they work through them \cite{anderson1993ProblemSolvingLearning}. 
% From an educational perspective, this suggests that novice learners might benefit from instruction that explicitly emphasizes the iterative nature of data science and encourages them to revisit earlier stages of the process as they gain new insights from the data. 
% It might also be helpful to provide scaffolding that supports novices in making decisions about when and how to iterate.}

Agglomerative Hierarchical Clustering (AHC) of the problem-solving sequences yielded an optimal solution of three clusters with an Average Silhouette Width = 0.28, indicating a discernible but weak cluster structure (\cite{saqr2024SequenceAnalysisEducation}). 
The clusters display distinct patterns (Figure \ref{fig:densityClusters}):
\begin{enumerate}
    \item Cluster 1 (n=173): Characterized by long sequences that generally follow the problem-solving stages in a linear order.
    \item Cluster 2 (n=242): Contains shorter sequences with an earlier and more prominent second peak of \textit{Data Collection} and earlier instances of \textit{Reflection}.
    \item Cluster 3 (n=25): A small cluster distinguished by a significantly higher proportion of \textit{Data Interpretation} actions.
\end{enumerate}
Addressing RQ2, a Chi-square test of independence revealed a significant association between cluster membership and expertise level ($\chi^2=11.85$ and $p=0.0027$).
Novices were overrepresented in Cluster 1 (Observed=148, Expected=133), while experts were overrepresented in Cluster 2 (O=68, E=55) and Cluster 3 (O=8, E=5).

\subsection{Phase-Level Transitions: Process Mining and Markov Models}
To examine the workflows with greater detail, we shift our analysis from the five high-level problem-solving stages to the 11 more granular DS phases.
This shift is motivated by the fact that initial analyses using the five broad stages yielded nearly identical process models for novices and experts for PM and MM. 
\begin{figure}
    \centering
    \includegraphics[width=0.9\linewidth]{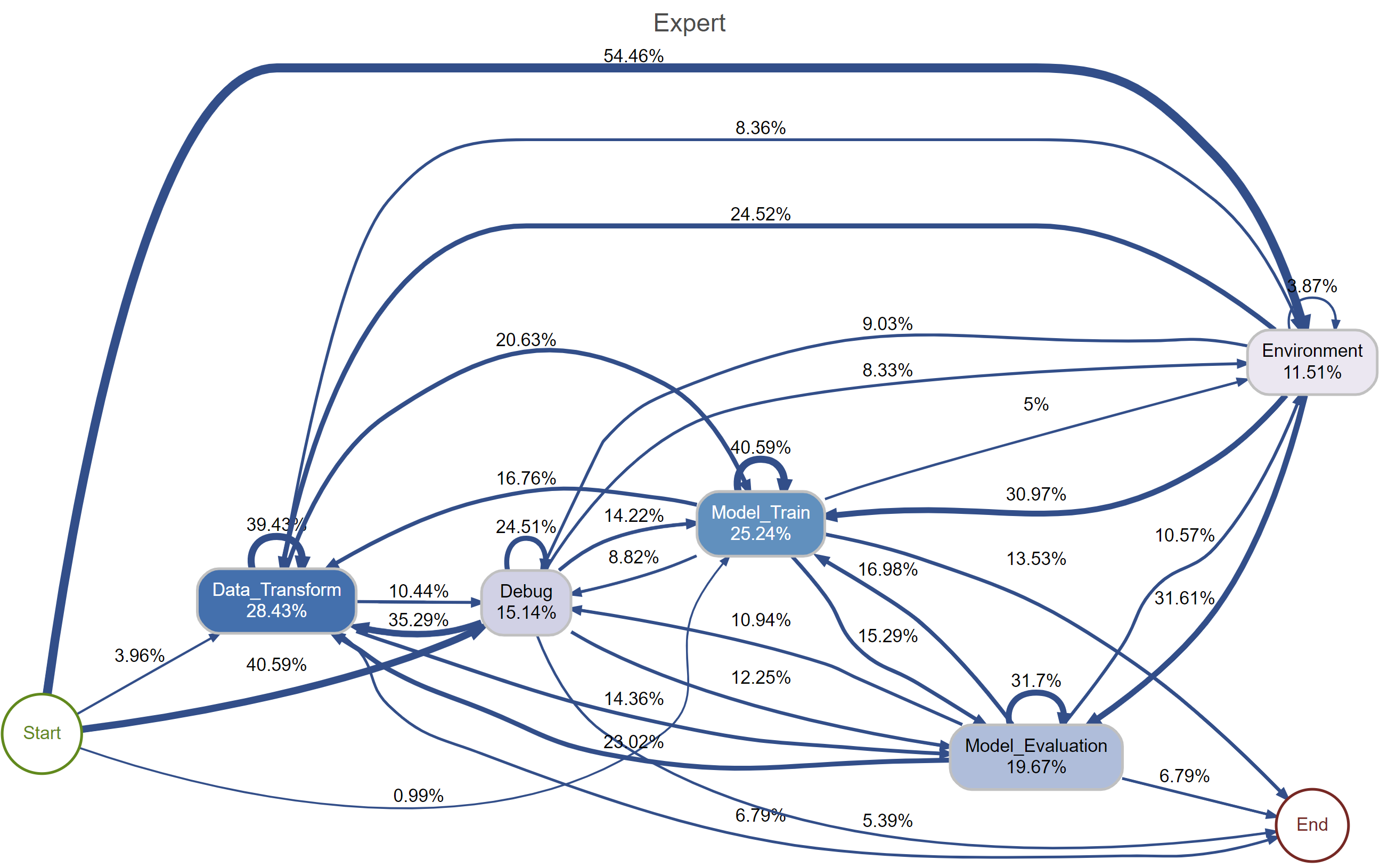}
    \caption{Process Visualisation for Experts}
    \label{fig:processMiningExpert}    
    \includegraphics[width=0.9\linewidth]{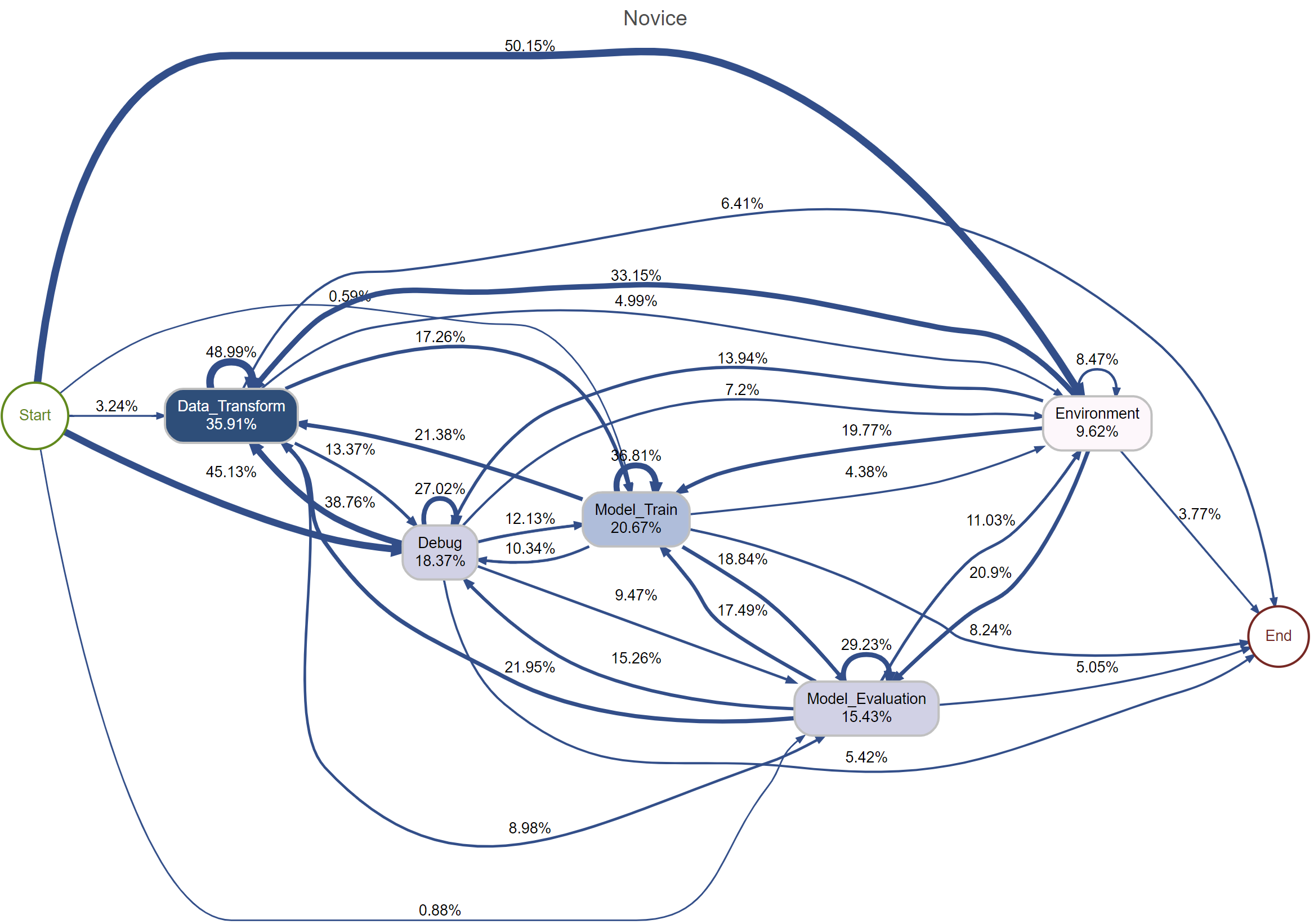}
    \caption{Process Visualisation for Novices}
     \label{fig:processMiningNovice}
\end{figure}

Process Mining (PM) models were generated for novice and expert groups at the DS phase level. 
The thickness of the arrows is proportional to the frequency of transitions between phases, and the percentages inside the nodes represent the proportion of all actions that occurred in that phase.
The resulting process maps (Figures \ref{fig:processMiningExpert} and \ref{fig:processMiningNovice}) identify largely similar workflows (RQ1).
However, subtle differences emerge in the transitions between novices and experts (RQ2). 
Novices transition to \textit{Data Transform} actions more frequently overall (35.9 vs 28.4\%), while experts show a slightly higher tendency to begin with \textit{Environment setup} (see the line from \textit{Start} to \textit{Environment}) and exhibit more frequent transitions involving \textit{Model Train} and \textit{Model Evaluation}.
 
% \textcolor{red}{The subtle differences observed in the process models suggest that novices might benefit from guidance on the importance of environment setup and planning before diving into data manipulation. 
% The higher frequency of data transformation actions among novices might indicate a trial-and-error approach, which could be addressed through instruction that emphasizes the importance of understanding the data and planning transformations more strategically.}
% Unfortunately, the differences between process models are not enough to identify particular differences in the strategies or to confirm the patterns discovered in the previous section.

A direct comparison of Markov Models (MM) built for novices and experts also revealed highly similar transition probabilities between DS phases. 
The most notable difference was a slightly higher probability for experts to transition from \textit{EDA} to \textit{Data Transform}. 
Given these subtle differences at the individual transition level, we proceeded with clustering to identify more holistic strategic patterns.

\begin{figure}
    \centering
    \includegraphics[width=0.6\textwidth]{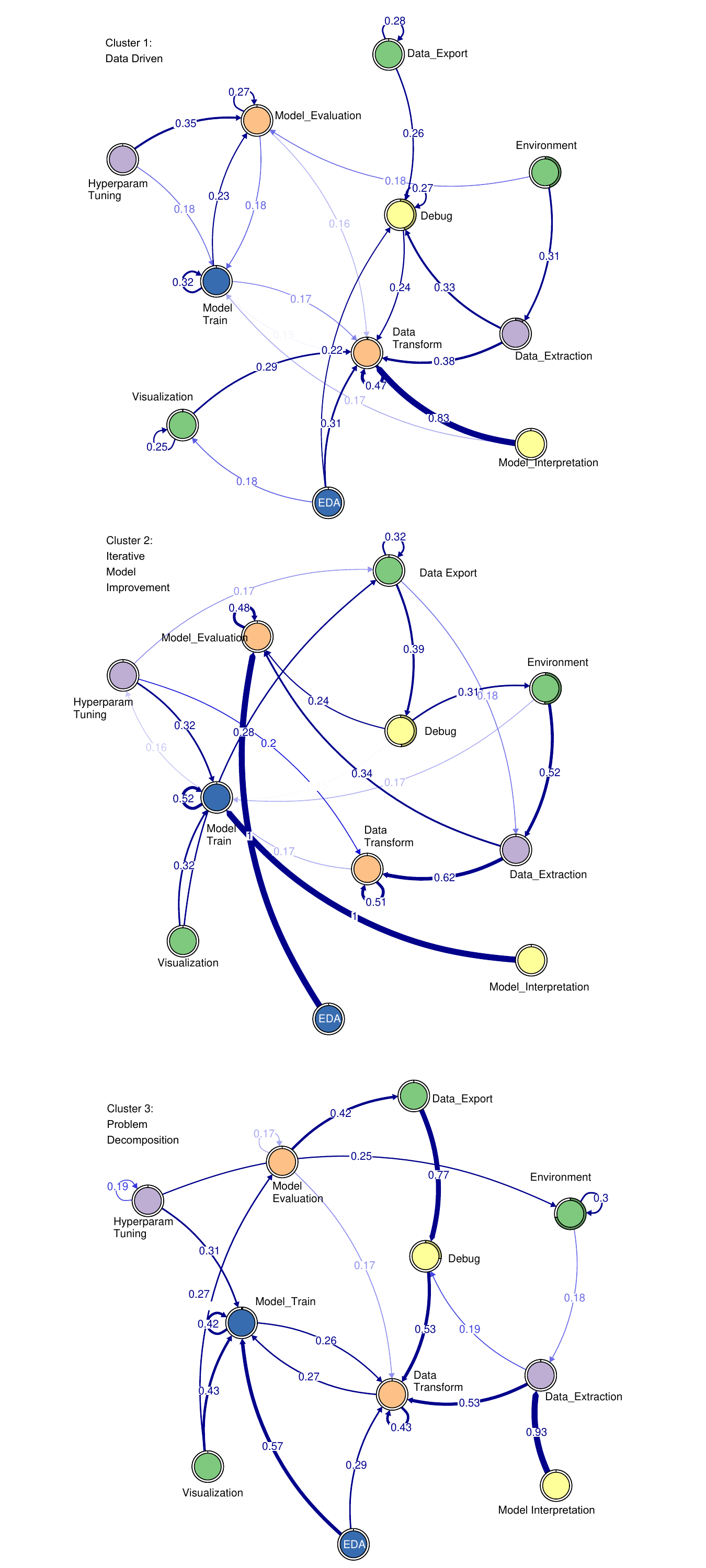}
    \caption{Transition Probabilities of Sequences by Cluster}
    \label{fig:markovClusters}
\end{figure}

The MM clustering analysis revealed three distinct strategic clusters based on their transition probabilities (RQ1) in Figure \ref{fig:markovClusters}. 
We label these based on their dominant patterns:
\begin{enumerate}
    \item A Data-driven cluster, characterized by frequent self-loops on \textit{Data Transform} and many incoming transitions to it.
    \item An Iterative Model Improvement cluster, defined by strong self-transitions (the thick, round arrows with 0.5) in \textit{Model Training} and \textit{Model Evaluation}, as well as from \textit{Model Interpretation} back to \textit{Model Training}.
    \item A Problem Decomposition cluster, which showed a high probability of starting (the circle around the node) with and repeating \textit{Environment} actions before moving deeper into the workflow.
\end{enumerate}
However, when testing if these strategies were associated with expertise (RQ2), a Chi-square test found no significant association between the MM-based 
clusters and expertise level ($\chi^2=0.053$ and $p=0.97$).

% \begin{figure}
%     \centering
%     \includegraphics[width=0.6\linewidth]{figures/MM_Class2.png}
%     \caption{Markov model of the Quora Classification Competition}
%     \label{fig:markovClass}
% \end{figure}

% In programming, we would expect expert students to follow such a top-down approach \cite{anderson1993ProblemSolvingLearning}.
% \textcolor{red}{The differences in transition probabilities between novices and experts, although subtle, suggest that experts might be more strategic in their transitions between data extraction, transformation, and EDA. 
% This could indicate a more integrated understanding of how these different stages of the data science process inform each other.  
% The higher transition probability from Model Training to Model Evaluation among experts suggests a greater emphasis on evaluating and refining models. 
% From an educational perspective, these findings suggest that instruction could focus on helping novices develop a more holistic understanding of the relationships between different data science activities. 
% For example, exercises could be designed to explicitly link data extraction decisions to downstream modelling choices, or to emphasize the iterative relationship between EDA and data transformation.}
% However, it's important to interpret these transition probabilities in the context of the overall data science workflow. 
% For example, the relatively high probability of transitioning from Environment to Data Extraction for both novices and experts reflects the common practice of setting up the environment before loading the data. 
% Similarly, the frequent transitions between Data Transform, EDA, and Model Training reflect the iterative nature of these activities in many data science projects. 

\subsection{Action Patterns Within Problem-Solving Stages}
For our final level of analysis, we return to the five high-level problem-solving stages to provide the context for discovering fine-grained action patterns. 
This approach allows us to identify which specific DS actions are frequently used in sequence to achieve a broader goal (e.g., Data Collection), rather than analyzing patterns within a single, more narrow DS phase. 
To identify these procedures (RQ1) and compare their usage by expertise (RQ2), we used Sequential Pattern Mining (SPM). 
Table \ref{tab:spm_patterns} summarizes notable patterns where frequencies differed between novices and experts.
Some notable examples are:
\begin{itemize}
    \item Problem Definition: Experts more frequently showed patterns involving repeated import\_modules actions. Novices, in contrast, uniquely showed a [load\_from\_csv, show\_table] pattern (7\% support).
    \item Data Collection: Novices frequently exhibited a [feature\_engineering, feature\_engineering] pattern (7\%). Expert patterns were dominated by model-related actions like [model\_class, fit] (6\%) and [model\_class, compile] (5\%).
    \item Data Recording: A [distribution, distribution] pattern was common among novices (15\%), while experts showed patterns like [model\_coefficients, learning\_history] (6\%).
    \item Reflection: Experts showed a [save\_to\_csv, save\_to\_csv] pattern (5\%) not present among novices.
\end{itemize}

\begin{table}
\footnotesize
\centering
\caption{Top Sequential Patterns by Problem-Solving Phase and Expertise}
\label{tab:spm_patterns}
\begin{tabular}{p{3cm}p{6cm}p{1.7cm}p{1.7cm}}
\toprule
Phase & Pattern & Novices & Experts \\
\midrule
Problem Definition and Decomposition & [import modules, load from csv] & 11\% & 17\% \\
& [list files, load from csv] & 10\% & 12\% \\
& [load from csv, show table] & 7\% & - \\
& [import modules, import modules] & - & 9\% \\
\midrule
Data Collection & [feature engineering, feature engineering] & 7\% & - \\
& [model class, fit] & - & 6\% \\
& [model class, compile] & - & 5\% \\
\midrule
Data Recording & [distribution, distribution] & 15\% & - \\
& [model coefficients, learning history] & - & 6\% \\
\midrule
Data Interpretation & Same patterns for both groups & - & - \\
\midrule
Reflection & [save to csv, save to csv] & - & 5\% \\
\bottomrule
\end{tabular}
\end{table}

% \textcolor{red}{The differences in sequential patterns between novices and experts provide further evidence for the idea that experts engage in more planning and preparation before diving into data manipulation, as well as a greater focus on model building and selection over the trial-and-error approach to feature engineering shown by novices.
% This suggests that novices might benefit from instruction that focuses on more principled approaches to feature engineering and model selection. 
% Overall, the SPM  findings can be used to inform the design of targeted interventions, such as providing scaffolding for planning and environment setup, guiding novices through more principled approaches to feature engineering and model selection, and emphasizing the importance of model evaluation and documentation.} 

\section{Discussion and Implications} %%%%%%%%%%%%%%%%%%%%%%%%%%%%%%%%%%%%%%
%%%%%%%%%%%%%%%%%%%%%%%%%%%%%%%%%%%%%%%%%%%%%%%%%%%%%%%%%%%\\
\label{sec:discussion}
This study used a multi-level sequence analysis to compare the problem-solving processes of novice and expert data scientists. 
Our central aim was to identify characteristic strategies that could differentiate expertise levels and, ultimately, inform data science education. 
Our findings confirm that discernible strategies exist (RQ1) and that the nature of these strategies depends heavily on the level of analytical granularity. 
While high-level phase transitions are similar across skill levels, the overall process structure and, most notably, the fine-grained action patterns within those processes reveal significant differences between novices and experts (RQ2).

\subsection{Interpreting Strategies Across Different Granularities}
By analyzing the notebooks at three distinct levels, we constructed a detailed picture of the differences in workflows.
At the overall process level, our analysis shows that experts and novices structure their entire workflow differently. 
Novices were significantly more likely to belong to a cluster characterized by a long, linear, step-by-step process, suggesting they adhere closely to a canonical workflow. 
In contrast, experts were overrepresented in a cluster featuring shorter, more iterative workflows. 
This aligns with classic literature on expertise, which finds that experts possess flexible mental schemas that allow them to adapt their strategies and work more efficiently (\cite{chi1981CategorizationRepresentationPhysics,anderson1993ProblemSolvingLearning}). 
However, the low silhouette score for this clustering (0.28) suggests that these ``strategies" are not rigid archetypes but rather overlapping tendencies, highlighting the inherent flexibility and variability in the data science process, even at different skill levels.

At the phase-transition level, the analysis revealed a crucial insight: the high-level order of operations is not a strong differentiator of expertise.
Our initial analysis using the five broad problem-solving stages showed no discernible differences between groups. 
Even when ``zooming in" to the 11 more granular DS phases, the non-significant result from the Markov Model clustering confirms that experts do not follow a fundamentally different sequence of phases than novices. 

At the action-pattern level, the procedures identified by SPM provide the concrete evidence that explains the high-level differences. 
The expert tendency toward shorter, iterative workflows (identified by AHC) is likely enacted through efficient modeling patterns like \texttt{[model\_class, fit]}. 
In contrast, the novices' linear, step-by-step approach is reflected in more exploratory patterns like \texttt{[feature\_engineering, feature\_engineering]} and \texttt{[load\_from\_csv, show\_table]}. 
These fine-grained patterns are the tangible evidence of the abstract strategies, explaining how experts achieve their efficiency and iterative nature.

\subsection{Implications for Data Science Education}
The patterns identified in this study have direct implications for designing learning experiences and support systems in data science.
\begin{itemize}
    \item Curriculum Design: The curriculum should explicitly teach and model the iterative nature of data science (\cite{donoghue2021TeachingCreativePractical}). This focus on process becomes even more critical in an era of Generative AI (\cite{ellis2023NewEraLearning}). As AI tools automate the generation of code snippets, the core human skill shifts from writing code to strategically guiding the problem-solving process—a skill our analysis identifies as a key component of expertise. Project-based learning activities that require students to cycle back through the workflow can help move them beyond a rigid, linear mindset (\cite{hazzan2020TenChallengesData}).
    \item Teaching Practices: Instructors can use these findings to provide targeted feedback. A notebook that shows a linear process with many trial-and-error transformations could prompt a discussion about planning and iterative refinement. Instructors can also use worked examples or live coding to model expert practices, explicitly narrating why they are returning to an earlier stage of the process (\cite{hazzan2023GuideTeachingData}).
    \item Navigating Generative AI: Our findings highlight a potential pitfall of Generative AI in education; its use can inadvertently reinforce the linear, non-iterative process characteristic of novices. A student who simply prompts an AI for the "next step" may get correct code but fails to develop the strategic, reflective, and iterative thinking that defines expertise. Educators must therefore focus on teaching students how to use these tools as a partner for brainstorming and refinement, not as a replacement for critical thought (\cite{fan_beware_2024}).
    \item Intelligent Support Systems: The patterns identified here can form the basis for automated feedback systems. These systems can be significantly enhanced by Generative AI. An AI-powered assistant, informed by our findings, could not only detect novice patterns but also provide expert-like scaffolding. For example, by prompting a student to reflect on model performance before attempting more feature engineering, it could actively encourage an expert-like iterative cycle (\cite{xu_enhancing_2025}).
    \item Assessment: This process-oriented analysis suggests new forms of assessment. Instead of only grading the final notebook or model accuracy, instructors could assess the process itself. The structure of a student's workflow, as captured by their sequence of actions, could become a valuable formative assessment tool to gauge their problem-solving maturity (\cite{molenaar_temporal_2022}).
\end{itemize}

\subsection{Limitations and Future Work} 
A primary limitation of this study is its intentional focus on a single, comprehensive Kaggle competition in text classification \cite{chakraborty2023QuoraInsincereQuestions}.
While this case study approach provides the depth necessary to analyze a wide spectrum of authentic procedures, the specific, low-level action patterns may not generalize to other contexts.
We contend, however, that the core practical implications for education are broadly transferable. 
The analytical framework presented here can be readily applied to other data science domains, such as image analysis or time-series forecasting, as well as to different scopes like short in-class work or scaffolded assignments, to investigate which problem-solving strategies are universal and which are task-specific.

A second limitation is that we analyze only the final state of notebooks, which misses the real-time process of editing, deleting, and debugging cells.
Future research should address these limitations by capturing more fine-grained, real-time interaction data from within notebook environments across a wider variety of tasks, learner expertise levels, and educational contexts (\cite{valle_torre_jelai_2025,cai_jupyter_2025}). 
Such data would enable a more dynamic analysis of the learning process, paving the way for adaptive support systems that can provide real-time (\cite{delaat2020ArtificialIntelligenceRealtime}), contextualized scaffolding to learners as they navigate the complexities of data science problem-solving (\cite{shen2024ImplicationsChatGPTData,zhao2023DataMakesBetter}).

\section{Conclusion}
This study set out to make the tacit problem-solving processes of data science visible, using a multi-level sequence analysis to compare the workflows of novices and experts. 
While our analysis of phase-to-phase transitions showed that novices and experts follow a similar order of operations, telling differences emerged at other levels of granularity. 
We found a statistically significant difference in the overall structure of their workflows, with novices tending toward longer, more linear processes, while experts demonstrated shorter, more iterative approaches. 
These high-level strategic differences were explained by the fine-grained action patterns within each phase, where experts employed more technically efficient procedures.

The central contribution of this work is the demonstration that data science expertise lies not in following a different or unique sequence of steps, but in how the process is enacted—with flexibility, iteration, and efficiency. 
This insight has critical implications for education, especially with the integration of Generative AI into data science tools and environments. 
Educational approaches that only teach a canonical, step-by-step workflow risk reinforcing the very novice-like behaviors we identified. 
Instead, assignments, materials, and intelligent support tools must be carefully designed to cultivate the adaptive, reflective, and iterative thinking that truly defines an expert practitioner, ensuring that new AI assistants become partners in developing expertise, not crutches that stifle it.

% \section{Data Availability Statement}
% Data sharing is not applicable to this article as no new data were created or analyzed in this study.

% \bigskip
% \begin{center}
% {\large\bf SUPPLEMENTARY MATERIAL}
% \end{center}

% \begin{description}

% \item[Title:] Brief description. (file type)

% \item[R-package for  MYNEW routine:] R-package ÒMYNEWÓ containing code to perform the diagnostic methods described in the article. The package also contains all datasets used as examples in the article. (GNU zipped tar file)

% \item[HIV data set:] Data set used in the illustration of MYNEW method in Section~ 3.2. (.txt file)

% \end{description}

\bibliographystyle{chicago}

\bibliography{references}

\end{document}